\documentclass[lettersize,journal]{IEEEtran}
\usepackage{amsmath,amsfonts}
\usepackage{algorithmic}
\usepackage{array}
\usepackage[caption=false,font=normalsize,labelfont=sf,textfont=sf]{subfig}

	\usepackage{cite}
\usepackage{siunitx}
\usepackage{amsmath,amssymb,amsfonts}
\usepackage{bm}
\usepackage{xcolor}
\usepackage[T1]{fontenc}
\usepackage{relsize}

\usepackage{textcomp}
\usepackage{stfloats}
\usepackage{url}
\usepackage{verbatim}
\usepackage{graphicx}
\def\BibTeX{{\rm B\kern-.05em{\sc i\kern-.025em b}\kern-.08em
		T\kern-.1667em\lower.7ex\hbox{E}\kern-.125emX}}
\usepackage{balance}

\begin{document}
	
	\title{Optimizing Network Performance and Resource Allocation in HAPS-UAV Integrated Sensing and Communication Systems for 6G\\
		
	}
	
\author{Parisa Kanani, Mohammad Javad Omidi, Mahmoud Modarres-Hashemi, and Halim Yanikomeroglu, Fellow, IEEE%
	\thanks{P. Kanani, M. J. Omidi, and M. Modarres-Hashemi are with the 
		Department of Electrical and Computer Engineering,
		 Isfahan University of Technology, Isfahan 84156-83111, Iran (emails: p.kanani@iut.ac.ir; omidi@iut.ac.ir; modarres@iut.ac.ir).}%
	\thanks{M. J. Omidi is also with the Department of Electronics and Communication Engineering, Kuwait College of Science and Technology, Doha 35003, Kuwait.}%
	\thanks{H. Yanikomeroglu is with Non-Terrestrial Networks (NTN) Lab, Department of Systems and Computer Engineering, Carleton University, Ottawa, ON K1S 5B6, Canada (email: halim.yanikomeroglu@sce.carleton.ca).}%
}

	
	\maketitle
	
	\begin{abstract}
		This paper proposes an innovative approach by leveraging uncrewed aerial vehicles (UAVs) as base stations (BSs) and a high-altitude platform station (HAPS) as the central processing unit (CPU) in an integrated sensing and communication (ISAC) system for 6G networks.
		We explore the challenges, applications, and advantages of ISAC systems in next-generation networks, highlighting the significance of optimizing position and power control. Our approach integrates HAPS and UAVs to enhance wireless coverage, particularly in remote areas. 
		UAVs function as dual-purpose access points (APs), using their maneuverability and line-of-sight (LoS) aerial-to-ground (A2G) links to transmit combined communication and sensing signals. The scheme operates in two time slots: in the first slot, 
		UAVs transmit dedicated signals to communication users (CUs)
		and potential targets. UAVs detect targets in specific ground locations and, after signal transmission, receive reflected signals from targets.
		In the second slot, UAVs relay these signals to HAPS, which performs beamforming to align signals for each CU from various UAVs. 
		UAVs decode information from HAPS and adjust transmissions to maximize the efficiency of the beam pattern toward the desired targets. We formulate a multi-objective optimization problem with the goal of maximizing both the minimum signal-to-interference-plus-noise ratio (SINR) for CUs and the echo signal power from the targets. This is achieved by finding the optimal power allocation for CUs in each UAV, subject to constraints on the maximum total power in each UAV and the transmitted beam pattern gain.
		Simulation results demonstrate the effectiveness of this approach in enhancing network performance, resource allocation, fairness, and system optimization. By utilizing HAPS as the CPU, computational tasks are offloaded from UAVs, which conserves energy and further improves overall network performance. 
		
	\end{abstract}
	\begin{IEEEkeywords}
		High altitude platform stations (HAPS), integrated sensing and communication (ISAC), optimization, sixth-generation (6G),   uncrewed  aerial vehicle (UAV), wireless networks
		
	\end{IEEEkeywords}
	
	\section{Introduction}
	Recently, integrated sensing and communication (ISAC) has emerged as a key technology for the next-generation communication systems, beyond the current fifth-generation (5G) and sixth-generation (6G) systems. In ISAC, sensing and communication systems are jointly designed to share frequency bands and hardware resources, resulting in improved energy efficiency and reduced hardware costs. The main objective of ISAC is to integrate sensing and communication processes into a cohesive system, resulting in mutual performance benefits \cite{iot,dual,limited,moti}.  This integration is expected to deliver notable enhancements in energy efficiency and spectral efficiency, while at the same time reducing hardware and signaling costs.
	
	Due to its ability to use common resources for both sensing and communication functions, such as hardware, waveforms, and frequency bands, ISAC has garnered significant attention in industry and academia. It is predicted that the use of ISAC will result in improved power consumption, reduced signal transmission delays, smaller product dimensions, and enhanced energy and spectral efficiency \cite{joint,dual,moti,limited}. Furthermore, the use of ISAC also leads to increased localization accuracy, suitable beamforming vector shaping, and reduced overhead for channel state information (CSI) tracking. 
	
	The use of ISAC in terrestrial networks is subject to notable constraints, especially in the context of sensing. The reason for this is twofold. Firstly, sensing tasks such as target detection or parameter estimation typically rely on direct links between the transmitter and targets. However, in terrestrial networks, obstacles in the environment can often obstruct direct line-of-sight (LoS) links, resulting in indirect paths that introduce errors and render some sensing operations infeasible. Secondly, achieving precise, long-range sensing demands a significant power input, potentially leading to a decline in performance for terrestrial base stations (BSs) \cite{dual,traj,man,thr}.
	
	Utilizing on-demand deployment and leveraging the LoS links offered by
	uncrewed
	aerial vehicles (UAVs) \cite{dual, moti}, UAVs are poised to emerge as highly promising aerial ISAC platforms. By catering to the specific requirements of sensing frequency and communication quality, UAVs have the potential to deliver more controlled and balanced integrated services. This is made possible through their capabilities in monitoring, sensing, and remote operations. Additionally, exploiting UAVs enables the attainment of high-resolution, all-weather, day-and-night imaging and high frame rates \cite{dual, iot, uav}.
	
	The paper \cite{cons} proposes optimizing multiple UAVs to act as communication providers and distributed MIMO radar for target sensing. The joint optimization of UAV location, communication user (CU)
	association, and transmission power control maximizes network utility while meeting localization accuracy constraints.
	
	The authors in \cite{man} discuss a UAV-based ISAC system where the UAV acts as an aerial access point for both communication and sensing purposes. The UAV sends combined information and sensing signals to communicate with multiple CUs while sensing potential targets on the ground. The sensing beampattern gain constraint is used as a sensing metric, and the weighted sum rate is used as a communication metric. This problem is similar to the one discussed in reference \cite{traj}.
	
	The paper \cite{energy} proposes an energy-efficient computation offloading strategy for a UAV-assisted edge computing system that uses ISAC. The proposed system model prioritizes sensed data and performs weighted allocation of computational resources on the ground roadside unit (RSU) based on the priority of vehicle perception data to minimize energy consumption and total CU latency.
	
	In recent years, the integration of high-altitude platform stations (HAPS) and UAV into communication networks has garnered significant attention as a very promising solution for expanding wireless network coverage and providing access to remote areas.
	HAPS has emerged as a notable technology in 6G communications \cite{vision}. Recognized by numerous organizations and research studies, non-terrestrial networks are regarded as a crucial and cost-effective element for establishing high-capacity connections in 6G wireless networks \cite{non}. The lack of adequate coverage in remote and hard-to-reach areas, including rural regions, poses significant challenges for current networks \cite{rur}.
	
	Moreover, even technologically advanced countries face issues with their existing telecommunications infrastructure, which lacks the necessary reliability to meet the demands of future-generation applications \cite{rur,super}. Additionally, these infrastructures prove highly vulnerable during natural disasters, with connectivity disruptions resulting in substantial property damage, business disruptions, and potential loss of life. As a result, strengthening ground communications through the integration with aerial networks like HAPS remains a key initiative in the development of 6G communication systems.
	
	Due to its high altitude, HAPS can provide continuous coverage, which can reduce the number of required cellular towers and consequently lead to a reduction in capital and operational costs. Furthermore, the mobility of drones enables dynamic deployment in high-density CU areas, thereby improving the overall network capacity.
	In \cite{super}, HAPS was proposed as a super macro base station for highly populated metropolitan areas. This idea was further supported in \cite{cach} as an extension of edge computing, and in \cite{comm} as an enabling technology for communication, computing, caching, and sensing in aerial delivery networks. In \cite{link}, the link budget of aerial platforms equipped with smart surfaces was analyzed and compared to terrestrial networks. HAPS is an excellent option for use as a CPU due to minimal blockage and shadowing in backhaul links with UAVs, ensuring reliable LoS links. It is particularly useful for backhauling aerial access points (APs) deployed to serve CUs in remote or poorly connected areas where terrestrial infrastructure may be lacking or damaged \cite{uxnb,haps6}.
	
	Results obtained from sensing are generally required for subsequent processes. A key challenge for UAVs performing sensing tasks is their limited computational capability and the need for low latency in data processing. For instance, processing all received local echo signals in UAVs can be highly time-consuming, and in latency-sensitive missions like target tracking in ISAC operations, unmet latency requirements can jeopardize the performance of ISAC \cite{uavka}.
	To address this issue, one feasible solution is offloading some computationally intensive sensing tasks, such as raw data or processed data, to a powerful central server.
	
	To achieve this, we utilized the HAPS as a powerful data center in this study to collect and integrate sensing data, resulting in improved accuracy and  richer information about the targets.
	According to our research, this is the first study that investigates the combination of HAPS and UAV with ISAC.
	This paper proposes using the sub-Terahertz (THz) frequency band (100 GHz to 300 GHz) for backhaul links between UAVs and HAPS, with the D band (110 - 170 GHz) as the carrier frequency. The D band is considered  an interesting range of frequencies for beyond-5G systems \cite{band,line}.
	
	In this paper, we investigate the use of a multiple-input multiple-output (MIMO) HAPS-UAV system-enabled ISAC model. The HAPS is used as an aerial CPU for backhauling and processing signals from UAVs.
	To ensure accuracy in sensing potential targets and enhance signal reception for multiple CUs, we utilize beamforming techniques by employing steering vectors of a uniform planar array (UPA) at the UAVs to align incoming signals. This approach approximates a fully digital communication beamformer while ensuring that individual radar beampattern gain requirements are met \cite{good_uav}.
	
	Our study aims to optimize the multi-objective problem by maximizing both the minimum signal-to-interference-plus-noise ratio (SINR) for ground  CUs and the echo signal power from targets, while satisfying the sensing and communication constraints.
	The sensing metric used in this problem is related to the requirements on the beampattern gain in the direction of targets, while the SINR serves as the performance metric for communication.
	To achieve our goal, we introduce the HAPS-UAV system-enabled ISAC model in Section II. In Section \ref{formulation}, we formulate a multi-objective optimization problem that seeks to maximize the minimum SINR for ground  CUs and the echo signal power from targets, while considering constraints on the sensing beampattern gain and the maximum total power in each UAV.
	In Section \ref{simi}, we present the simulation results and methodology evaluation, showcasing the effectiveness of our proposed scheme. Finally, we present our discussion and conclusions based on these results in Sections \ref{discusiion} and \ref{natije}, respectively.
	
	%
	%
	\section{system model}
	\begin{figure}[t]
		\centering
		\includegraphics[width=0.47\textwidth]{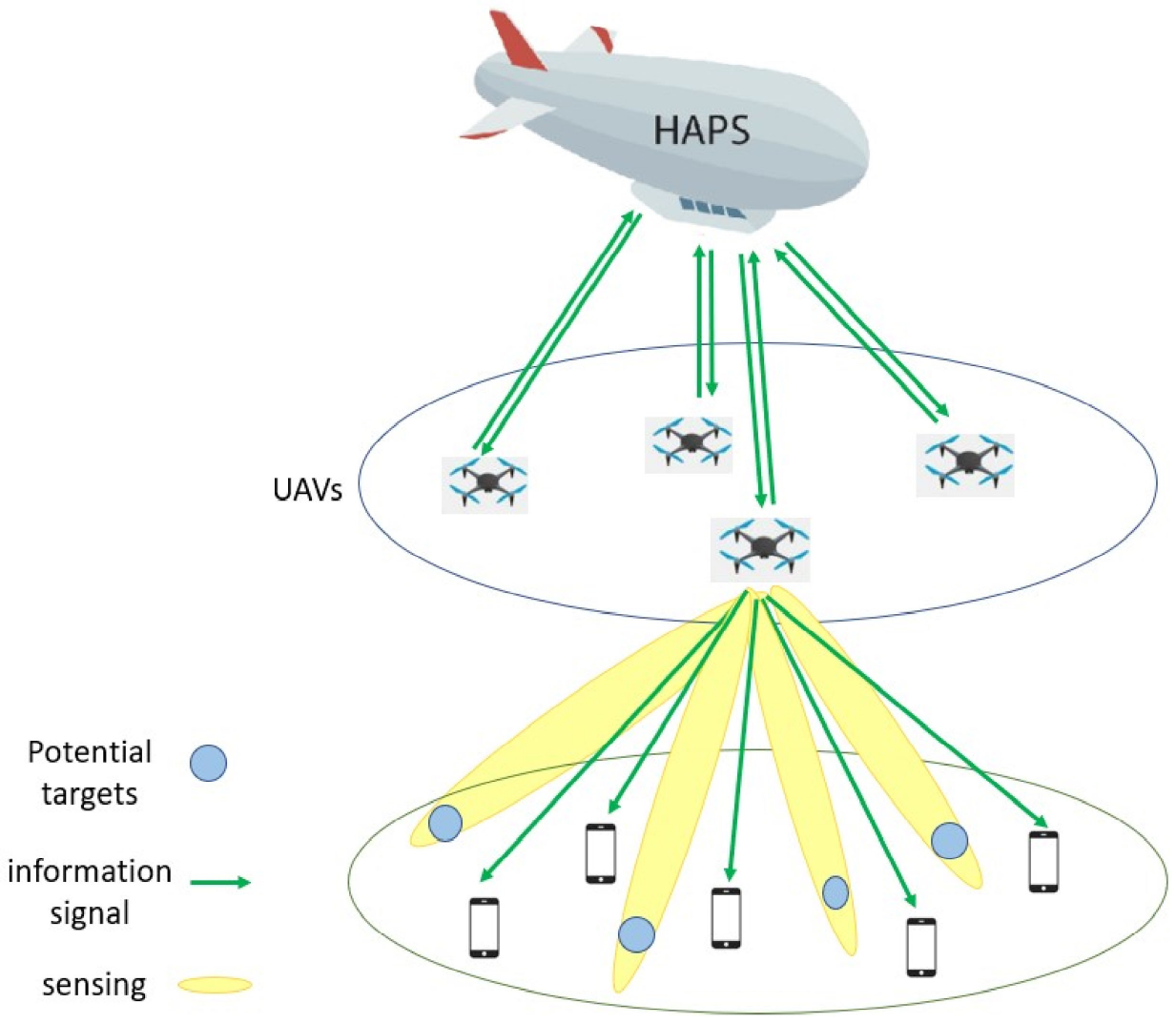}
		\caption{Illustration of a HAPS-UAV system-enabled integrated sensing and
			communication with downlink communication services for $K$ communication users
			and sensing capabilities for several potential ground targets.}
		\label{systemm}
	\end{figure}
	As shown in Fig. \ref{systemm}, the proposed model integrates HAPS and UAVs, where the UAVs sense specific ground targets while also facilitating communication between ground CUs.
	Each UAV is assumed to be assigned to specific ground CUs and designated targets. The objective of sensing designated ground points is to enable the detection of targets at those precise locations \cite{thr,man}. 
	The assumption is that CUs are single-antenna devices, while the UAVs are equipped with a UPA consisting of $G$ antennas, where $G = G_w \times G_l$, with $G_w$ and $G_l$ representing the number of antennas along the x-axis and the y-axis, respectively.  The total count of UAVs is denoted by $M$, and each UAV is uniquely identified by the index $m$, where $m$ belongs to the set $\mathbb{M} = \left\{1, 2, \ldots, M\right\}$.
	Also, the HAPS is equipped with a UPA containing a large number of antenna elements ($S=S_w \times S_l$), where $S_w$ and $S_l$ denote the number of antenna elements in the width and length of the array, respectively.
	
	In this model, the number of ground CUs assigned to each UAV is $K$, and the number of targets assigned for sensing to each UAV is $J$.
	During the first time slot, the dedicated signal $s_k^m [n]$ is transmitted by the $m$-th UAV towards CU $k$ (where $k \in \mathbb{K} \triangleq \{1, 2, \ldots, K\}$),
	and the beamforming vector for communication with CU $k$ from UAV $m$ at time slot $n$ is denoted by $\mathbf{w}_k^m [n] \in \mathbb{C}^{G \times 1}$.
	Additionally, the $m$-th UAV transmits the dedicated signal ${s'_j}^m [n]$ towards the target $j$ (where $j \in \mathbb{J} \triangleq \{1, 2, \ldots, J\}$) simultaneously.
	These signals, i.e., $s_k^m [n]$ and ${s'_j}^m [n]$, are uncorrelated and independent random variables with a mean of zero and a variance of one for any time slot $n$.
	Here, $n \in \mathbb{N} \triangleq \{1, \ldots, N\}$ represents a discrete time slot. An ISAC period is denoted as $\tau \triangleq [0, T]$, which is divided into $N$ discrete time slots with a duration of each time slot $\Delta_t = \frac{T}{N}$. $N$ is chosen such that $\Delta_t$ is sufficiently small to assume the UAV's position to be approximately constant within each time slot.
	Based on this, the signal transmitted by the $m$-th UAV  for communication and sensing purposes in time slot $n$ can be expressed as
	\begin{equation}
		\mathbf{x}_m[n] = \sum_{k=1}^{K} \mathbf{w}_k^m [n] s_k^m [n] + \sum_{j=1}^{J} \mathbf{r}_j^m [n] {s'_j}^m [n] , \quad \forall n \in \mathbb{N},
		\label{y}
	\end{equation}
	in which $\mathbf{r}_j^m [n] \in \mathbb{C}^{G \times 1}$ represents the beamforming signal sent by UAV $m$ for sensing purposes.
	Furthermore, the average transmitted power from the $m$th UAV in time slot $n$ is given by $\mathbb{E}(\|\mathbf{x}_m[n]\|^2) = \sum_{k=1}^{K} \|\mathbf{w}_k^m[n]\|^2 + \sum_{j=1}^{J} \|\mathbf{r}_j^m[n]\|^2$.
	The UAV operates under a maximum transmit power constraint denoted as 
	$P^m_{\max}$, as illustrated in the following equation:
	\begin{equation}
		\sum_{k=1}^{K} \| \mathbf{w}_k^m [n] \|^2 + \sum_{j=1}^{J} \| \mathbf{r}_j^m [n] \|^2  \le P^m_{\max}, \quad \forall n \in \mathbb{N}
		\label{qeid}.
	\end{equation}

	The time-varying position of the UAV $m$ in time slot $n \in \mathbb{N}$ is $(x^m[n], y^m[n], H_m)$ and $\mathbf{q}^m[n] = (x^m[n], y^m[n])$ represents the horizontal flight location. Also, $H_m$ is the constant flight altitude. The horizontal location of the CUs is denoted as $\mathbf{u}_k^m [n] = (x_k^m [n], y_k^m [n])$, which can be determined using global navigation satellite system (GNSS) or estimation of the received signals \cite{man}. Additionally, the index $m$ in $\mathbf{u}_k^m$ denotes CUs associated with UAV $m$.
	Due to the high altitude of the UAV, there typically exists a strong line-of-sight (LoS) connection between the UAV and CU. Therefore, the wireless channel between UAV $m$ and CU
	 $k$ during time slot $n$, assuming the reciprocity of the downlink (DL) and uplink (UL) channels,  can be modeled as \cite{beam}
		\begin{IEEEeqnarray}{rCl}
		\mathbf{h}_k^m(\mathbf{q}^m[n], \mathbf{u}_k^m) 	&=&\mathbf{h}_{k,n}^{m, \text{DL}} 
		= \mathbf{h}_{k,n}^{m, \text{UL}} \\
		&=& \sqrt{\frac{\beta_{0_m}}{(d_{k,n}^m)^2}} \mathbf{a}_m(\mathbf{q}^m[n], \mathbf{u}_k^m). \nonumber
	\end{IEEEeqnarray}
	In this context,
	\begin{IEEEeqnarray}{rCl}
		d_{k,n}^m=d^m(\mathbf{q}^m[n], \mathbf{u}_k^m) = \sqrt{H_m^2+\lVert \mathbf{q}^m[n] - \mathbf{u}_k^m \rVert^2}
	\end{IEEEeqnarray}
 represents the distance between the UAV $m$ and CU $k$ in time slot $n$. Also, $\beta_{0_m}$ represents the channel power gain at a reference distance $d_0 = 1$\,\text{m}  and the steering vector $\mathbf{a}_m(\mathbf{q}^m[n],\mathbf{u}_k^m)$ towards CU $k$ is given by \cite{userg,graph,Multiuser,favorable}
	\begin{align}
		\mathbf{a}_m(\mathbf{q}^m[n], \mathbf{u}_k^m) = \boldsymbol{\alpha}_m(\theta_k[n],\phi_k[n]) \otimes \boldsymbol{\xi}_m(\theta_k[n],\phi_k[n]),
		\label{steer}
	\end{align}
\begin{equation}
	\begin{split}
		\boldsymbol{\alpha}_m (\theta_k[n],\phi_k[n]) & = 
		\begin{bmatrix}
			1, \  e^{-j2\pi(d\sin\theta_k[n]\cos\phi_k[n])/\lambda},  \cdots , \\
			e^{-j2\pi(G_w-1)(d\sin\theta_k[n]\cos\phi_k[n])/\lambda}
		\end{bmatrix}^T,
	\end{split}
\end{equation}
\begin{equation}
	\begin{split}
		\boldsymbol{\xi}_m(\theta_k[n],\phi_k[n]) & = 
		\begin{bmatrix}
			1, \ e^{-j2\pi(d\sin\theta_k[n]\sin\phi_k[n])/\lambda}, \ \cdots, \\
			e^{-j2\pi(G_l-1)(d\sin\theta_k[n]\sin\phi_k[n])/\lambda}
		\end{bmatrix}^T.
	\end{split}
\end{equation}
	In \eqref{steer}, $ \theta_k \in [0, \frac{\pi}{2}] $ and $ \varphi_k \in [-\pi, \pi] $  represent the vertical and the horizontal angle of departure (AoD) of the $k$-th CU,
	respectively. The symbol $\otimes$ is the Kronecker product. 
	$\lambda$ is the carrier wavelength, and $d$ is the distance between adjacent antennas. Consequently, the received signal at CU 
	$k$
	in time slot 
	$n$
	associated with UAV 
	$m$
	can be expressed as
	\begin{equation}
		z_k^m[n] = \mathbf{h}_k^H(\mathbf{q}^m[n], \mathbf{u}_k^m) \mathbf{x}_m[n] + v_k^m[n],
	\end{equation}
	where $v_k^m$ is the additive white gaussian noise (AWGN) with variance $\sigma_{k,m}^2$.
	The expression $\mathbf{h}_k^H(\mathbf{q}^m[n], \mathbf{u}_k^m)$ represents the Hermitian of $\mathbf{h}_k^m(\mathbf{q}^m[n], \mathbf{u}_k^m)$. Consequently, the SINR at CU $k$ can be formulated as follows:

	\begin{align}
		\text{SINR}_k=	\gamma_k^m[n] =
		\frac{{\left| \mathbf{h}_k^H(\mathbf{q}^m[n], \mathbf{u}_k^m) \mathbf{w}_k^m [n] \right|^2}} {  I_k^m[n]+ \sigma_{k,m}^2},
	\end{align}
	where 
	\begin{align}
		& I_k^m[n]=\\
		&\left| \sum_{i=1, i\neq k}^{K} \mathbf{h}_k^H(\mathbf{q}^m[n], \mathbf{u}_k^m) \mathbf{w}_i^m[n]  +\sum_{j=1}^{J}  \mathbf{h}_k^H(\mathbf{q}^m[n], \mathbf{u}_k^m) \mathbf{r}_j^m [n] \right|^2.\nonumber
	\end{align}
	In this case, the achievable rate at CU $k$ in time slot $n$ is given by $R_k[n] = \log_2(1 + \gamma_k^m[n])$.		 
	
	Drawing from references \cite{man} and \cite{thr}  in the sensing context, the UAVs aim to detect potential targets at a limited number of locations, J, on the ground. The horizontal positions of these targets are represented by
	$\mathbf{m}_j$ for $j \in \mathbf{J} \triangleq \{1, \ldots, J\}$. The values of $\mathbf{m}_j$ are determined based on specific sensing tasks of the UAVs \cite{thr}.
	Essentially, $\mathbf{m}_j$ represents the likely positions of these targets to aid in tracking.
	Upon transmitting signals from the UAVs for communication and sensing, the reflected signals from the targets are received by the UAVs. In this process, one of the desired outcomes, similar to \cite{man} and \cite{thr}, is to maximize the utilization of the transmitted beam pattern from the UAVs directed at location $\mathbf{m}_j$, represented by
	\begin{align}
		\zeta(\mathbf{q}^m[n], \mathbf{m}_j) & = \mathbb{E}\left[ \left| \mathbf{a}_m^H(\mathbf{q}^m[n], \mathbf{m}_j) \mathbf{x}_m[n] \right|^2 \right]  \\
		& = \mathbf{a}_m^H(\mathbf{q}^m[n], \mathbf{m}_j) \bigg( \sum_{k=1}^{K} \mathbf{w}_k^m[n] (\mathbf{w}_k^m[n])^H \nonumber \\
		& + \sum_{j=1}^{J} \mathbf{r}_j^m[n] (\mathbf{r}_j^m[n])^H \bigg) \mathbf{a}_m(\mathbf{q}^m[n], \mathbf{m}_j),\nonumber
		\label{beam}
	\end{align}
	and is intended to surpass a predetermined threshold. In the second slot, the UAVs relay the reflected signals from the targets to the HAPS when the received signal surpasses a predetermined  threshold. The communication link between the UAVs and HAPS utilizes the terahertz (THz) band \cite{cell}.
	We introduce a straightforward method to enhance the UAVs' sensing performance. This approach involves leveraging the known DL communication signals and the decoded CU
	signals to isolate and focus on the sensing signals from the received data, effectively excluding non-relevant components \cite{book_mimo ,limited}.
	Assuming successful distinction of echoes from different targets, we can express the echo 
	$\mathbf{y}_{m}[n]$	reflected from the targets and received by the 
	$m$-th UAV (with negligible consideration of the Doppler effect) using the equation described in \cite{limited}:

	\begin{IEEEeqnarray}{rCl}
	\mathbf{y}_{m}[n] &=& \sum_{j=1}^{J} \epsilon_{m,n}^j 
\mathbf{r}_j^m [n] \mathbf{a}_m^H(\mathbf{q}^m[n], \mathbf{m}_j) \label{echo} \\
\quad &\times& \mathbf{a}_m(\mathbf{q}^m[n], \mathbf{m}_j) s'_j [n-\tau_{m,n}^j ], \nonumber
	\end{IEEEeqnarray}
	
	\noindent where $\tau_{m,n}^j$, $\epsilon_{m,n}^j$ represent the 
	time delay and reflection coefficients, respectively,
	corresponding to
	cycle $n$ from the $j$-th target to the $m$-th UAV \cite{limited,beam}. 	 Our approach involves separating the desired sensing signals from the received signals by subtracting unwanted signals. Also, the steering vector 
	$\mathbf{a}_m(\mathbf{q}^m[n], \mathbf{m}_j)$
	can be obtained from equation \eqref{steer}, such that $\mathbf{m}_j$ is the horizontal location of the targets.
	To calculate \(\mathbf{y}_{m}[n]\), substitute the expression for $\mathbf{a}_m(\mathbf{q}^m[n], \mathbf{m}_j)$ into equation \eqref{echo}:
%
%
%
\begin{align}
	\mathbf{y}_{m}[n] &= \sum_{j=1}^{J} \epsilon_{m,n}^j \mathbf{r}_j^m [n] \\
	&\times \left(\boldsymbol{\alpha}_m(\theta_k[n],\phi_k[n]) \otimes \boldsymbol{\xi}_m(\theta_k[n],\phi_k[n])\right)^H \nonumber\\
	&\times \left(\boldsymbol{\alpha}_m(\theta_k[n],\phi_k[n]) \otimes \boldsymbol{\xi}_m(\theta_k[n],\phi_k[n])\right) s'_j [n-\tau_{m,n}^j ].\nonumber 
\end{align}

	In the second slot, the UAVs send reflected echo signals to the HAPS which serves as the central processing unit (CPU) for backhauling aerial UAVs for the HAPS-UAV system-enabled ISAC. The HAPS receiver processes the message of every UAV received through analog beamforming. Our goal is to determine the optimal power allocation values for each target within every UAV based on our analysis, ensuring the maximum beam pattern gain from each UAV directed toward its respective targets. 
	Analog beamforming is 
	utilized to steer the signal transmitted
	from each UAV towards the HAPS assuming direct LoS communication between them. The signal received from the UAVs by the antenna element $s$ of the HAPS is expressed as
	\begin{align}
		\mathbf{y'}_{s}[n] &= \sum_{m=1}^{M} \sum_{g=1}^{G} g_{mgs} b_{mg} \mathbf{y}_{m}[n] + Z_H[n] \nonumber \\
		&= \sum_{m=1}^{M} \sum_{g=1}^{G} c_{ms} \delta_{m} b_{mg} b_{mg}^{*} \mathbf{y}_{m}[n] + Z_H[n] \nonumber \\
		&= G \sum_{m=1}^{M} c_{ms} \delta_{m} \mathbf{y}_{m}[n] + Z_H[n],
		\label{haps}
	\end{align}
	where $Z_H[n]$ represents the AWGN at each receiving antenna element $s$ of the HAPS  in time slot $n$. Also,
	$g_{mgs}$ is
	the channel gain between the transmit antenna element $\mathbf{g}=(g_w, g_l)$ of UAV $m$ and the HAPS receiver antenna element
	$\mathbf{s}=(s_w, s_l)$  in the sub-THz frequency band and is assumed to be LoS. We assume that $g_{mgs} = \delta_{m} b^*_{mg}  c_{ms}$, 
	where $b_{mg}$ represents the phase shift of the transmitted signal from antenna element $g$ of UAV $m$ and $c_{ms}$ denotes the phase shift of the received signal from antenna element $s$ of the HAPS, as described in  \cite{uxnb}. Moreover, the term  $\delta^2_m$ represents the path loss between UAV $m$ and the HAPS, which has been addressed by \cite{uxnb}. The relation is expressed as
	
	\begin{align}
		b_{mg} &= \exp(j2\pi (\frac{d_m}{\lambda}) \times \exp(j\pi (g_w-1)
		\sin\Theta_m\cos\Phi_m)\nonumber\\
		&\quad\times\exp(j\pi (g_l-1)
		\sin\Theta_m\sin\Phi_m)),
		\label{bmg}
	\end{align}
	\begin{align}
		c_{ms} &= \exp(j\pi (s_w-1)
		\sin\Theta_m\cos\Phi_m)\nonumber\\
		&\quad\times\exp(j\pi (s_l-1)
		\sin\Theta_m\sin\Phi_m)
		\label{c}.
	\end{align}
	Here, $\Theta_m$ and $\Phi_m$ represent the elevation and azimuth angles, respectively, of the signal transmitted from UAV $m$ to the HAPS. Additionally, $d_m$ denotes the distance between the reference antenna element of UAV $m$ and the reference antenna element of the HAPS.
	Similar to the approach in \cite{cell}, we employ analog beamforming with phase shifters (PSs) to precisely steer transmitted signals from each UAV towards the HAPS. This procedure entails multiplying the signal by $b_{mg}$, as defined in \eqref{bmg}, for every transmission antenna element $g$ of each UAV $m$.
	The steering vector $\mathbf{b}_m = [b_{m,1},\dots,b_{m,G}]^T$ is responsible for the transmit antenna array of UAV $m$, while the steering vector $\mathbf{c}_m = [c_{m,1},\dots,c_{m,S}]$ corresponds to the receive antenna array of the HAPS, as transmitted from UAV $m$ \cite{cell,uxnb}. 
	The receiver design proposed for the HAPS employs 
	analog beamforming using phase shifters to precisely align the received signals from each UAV $m$ with the HAPS's receiving antennas. To achieve this, we multiply the signal $\mathbf{y'}_s[n]$ by the conjugate of the steering vector of the receive antenna elements at the HAPS for each UAV, which is represented as $c_{ms}^*$ in equation \eqref{c}, and then combine these signals as follows \cite{cell}:
	\begin{equation}
		\mathbf{y}[n] = \sum_{m=1}^{M} \sum_{s=1}^{S} c_{ms}^* \mathbf{y'}_s[n].
		\label{final}
	\end{equation}
	
	\section{Problem Formulation}
	\label{formulation}
	In this Section, we present the optimization problems for both sensing and communication within the HAPS-UAV-enabled ISAC system. The subsequent Subsections offer a comprehensive mathematical formulation for each objective, incorporating the relevant constraints and unique considerations associated with these functions. Finally, we integrate these objectives into a multi-objective optimization problem that harmonizes the sensing and communication goals, addressing the inherent trade-offs to achieve balanced and efficient performance across the entire system.

	
	\subsection {Optimizing HAPS-Received Signal Power to Enhance Sensing Performance in the ISAC System}
	To enhance  the  beam pattern gain of the transmissions 
	directed at
	the targets by UAVs, as outlined in \eqref{beam}, we recognize that  a higher  beam pattern gain in the signal  transmitted to the targets leads to increased received signal power in the UAV's receiver during 
	the targets' echo return.
	This, in turn, results in a higher received signal power by the HAPS. Indeed, maximizing this sensing metric is equivalent to maximizing the signal power of the targets. To achieve this objective, we will address the subsequent optimization problem:
\begin{align}
	\label{behine1}
	\max_{\mathbf{q}^m[n], \mathbf{r}_j^m[n] \,\, \forall j \in \mathbb{J}} & \quad \Omega \\
	\text{s.t.} \quad\quad\:\ & \hspace{-5mm} \sum_{j=1}^{J} \lVert \mathbf{r}_j^m[n] \rVert^2 \leq \upsilon_m P^m_{\max}, \tag{a} \\
	& \hspace{-5mm} \lVert \mathbf{q}^m[n+1] - \mathbf{q}^m[n] \rVert \leq V^m_{\max} \Delta t, \: \forall n \in \mathbb{N}, \tag{b} \\
	& \hspace{-5mm} \mathbf{q}^m_{\text{min}} \leq \mathbf{q}^m[n] \leq \mathbf{q}^m_{\text{max}}, \tag{c}
\end{align}
	where
	$	\Omega$ represents the signal power received at the HAPS, and it is determined by the following expression, as indicated by  \eqref{haps}, \eqref{final}, and \eqref{echo}: 
	\begin{equation}
		\label{omega1}	
		\Omega	=\left\vert \sum_{m=1}^{M} \sum_{s=1}^{S} c_{ms}^*G\sum_{m'=1}^{M}c_{m's}\delta_{m'} \mathbf{y}_{m'}[n] \right\vert^2.
	\end{equation}
	In this section, we've primarily concentrated
	on maximizing the received signal power, which originates from the echo of the
	signal transmitted from targets to the HAPS via the UAVs.
	Achieving this objective requires a coordinated optimization approach involving both the UAVs' transmitted beamforming vectors aimed at the targets, and the determination of optimal UAV positioning relative to the HAPS.
	Constrained by the maximum transmit power condition defined in constraint (\ref{behine1}.a), and characterized by the coefficient $\upsilon_m$ representing the proportion of $P^m_{\max}$ allocated by each UAV $m$ to its associated targets, the upper limit of transmission power for each UAV is fixed at $P^m_{\max}$. This power allocation encompasses a portion designated for the targets and another segment reserved for CU transmission. The power allocated to a specific set of targets linked with UAV $m$ is denoted as $\upsilon_m P^m_{\max}$.
	
	The constraint (\ref{behine1}.b), which represents the velocity constraint, signifies that the separation between the current position of the UAV ($\mathbf{q}^m[n]$) and its predicted position in the subsequent time step ($\mathbf{q}^m[n+1]$) must not surpass $V^m_{\text{max}} \Delta t$. 
	This velocity constraint indicates that the UAV moves at a constant speed during the specified time interval.
	Furthermore, the constraints [$\mathbf{q}^m_{\text{min}}$, $\mathbf{q}^m_{\text{max}}$] in (\ref{behine1}.c) are enforced to guarantee that the UAV's trajectory stays within a predefined range, ensuring regulated and controlled movement.
	Specifically, this means that the x-component of UAV $m$'s position, denoted as $q^m_{x}$, remains between $q^m_{x_{\text{MIN}}}$ and $q^m_{x_{\text{MAX}}}$, while its y-component, denoted as $q^m_{y}$, is similarly constrained.
	
	Another key objective of our study is to enhance the SINR for ground-based CUs during communication, while also ensuring adherence to the sensing requirements. As such, in the context of the proposed HAPS-UAV system-enabled ISAC, we significantly bolster the equity of SINR across  CUs. To achieve this aim, we have formulated an optimization problem, which is
	expounded upon in the following Subsection.
	\subsection{Optimizing CU SINR for Enhanced Communication Performance in ISAC System}
	\label{sub1}
	As mentioned earlier, beamforming in the HAPS is performed towards all UAVs to align the received signals for each CU from different UAVs. The UAVs utilize the information sent by the HAPS to adjust the transmit power towards the targets, maximizing the utilization of the transmitted beam pattern towards location $j$. An ISAC cycle encompasses all stages, including transmitting information signals $s_k[n]$ to CUs, transmitting signal $s'_j	[n]$ towards designated ground points by the UAVs, receiving the signals at the HAPS, which are the transmitted signals from the UAVs and include the echoes from the reflective targets' signals received by the UAVs, and adjusting the transmit power of the UAVs.
	
	The optimization problem, as defined in \eqref{mfi}, is formulated to maximize the minimum SINR for CUs. This involves determining the beamforming vector $\mathbf{w}_k^m[n]$ for the signal dedicated to CU,
	the transmit beamforming vector $\mathbf{r}_j^m[n]$  used in sensing, and $\mathbf{q}^m[n]$, representing the current location of UAV $m$. The optimization process seeks to find optimal values for these variables, ensuring the fulfillment of sensing requirements.
	%
	Accordingly, the optimization
	problem is formulated as
\begin{align}
	\max_{\mathbf{w}_k^m [n], \mathbf{q}^m[n], \mathbf{r}_j^m [n]} & \quad \min_k \quad \text{SINR}_k \label{mfi} \\
	\text{s.t.} \quad\quad\quad\quad\quad\:
	& \hspace{-4.4em} \sum_{k=1}^{K} \|\mathbf{w}_k^m[n]\|^2 + \sum_{j=1}^{J} \|\mathbf{r}_j^m[n]\|^2 \leq P^m_{\max}, \quad \forall n, j, m, k, \tag{a} \\
	& \hspace{-4.4em} \mathbf{q}^m[n+1] - \mathbf{q}^m[n]\| \leq V^m_{\max} \Delta t, \quad \forall n,m \tag{b} \\
	& \hspace{-4.4em} \mathbf{q}^m_{\text{min}} \leq \mathbf{q}^m[n] \leq \mathbf{q}^m_{\text{max}}, \quad \forall n, m \tag{c} \\
	& \hspace{-4.4em} \zeta(\mathbf{q}^m[n], \mathbf{m}_j) \geq d^2(\mathbf{q}^m[n], \mathbf{m}_j) \Gamma_j^{\text{th}}, \quad \forall j, n, m. \tag{d}
\end{align}
	The constraint on transmitted power is delineated in (\ref{mfi}.a), where $P^m_{\max}$ signifies the maximum power that UAV $m$ can transmit. This constraint not only imposes a limit on the transmitted power of each UAV ($m$), denoted by $P^m_{\max}$, but also ensures that the UAV operates within safe and acceptable power levels. The sensing metric is intertwined with constraints on the beam pattern gain toward targets, expressed in (\ref{mfi}.d).
	Moreover, constraint (\ref{mfi}.d) pertains to the beam pattern gain of UAV transmissions toward designated targets, as defined in equation \eqref{beam}. In this constraint, $\Gamma_j^{th}$ denotes the beam pattern gain threshold for target $j$, and $d^2(\mathbf{q}^m[n], \mathbf{m}_j)$ represents the corresponding path loss.
	%
	%

	We can reformulate the problem \eqref{mfi} by introducing an auxiliary variable \(\eta\) to simplify the optimization process. 
	To transform this into a more standard optimization problem, we introduce an auxiliary variable \(\eta\) that represents the minimum SINR we aim to maximize. The problem then becomes
\begin{align}
	\max_{\mathbf{w}_k^m [n], \mathbf{q}^m[n], \mathbf{r}_j^m [n], \eta \,\, \forall j \in \mathbb{J}, k \in \mathbb{K}} & \quad \eta \label{mfi2} \\
	\text{s.t.} \quad \quad\quad\quad\quad\quad\quad
	& \hspace{-5em} \begin{array}{@{}l@{}}
		\displaystyle \sum_{k=1}^{K} \|\mathbf{w}_k^m[n]\|^2 + \displaystyle \sum_{j=1}^{J} \|\mathbf{r}_j^m[n]\|^2 \leq P^m_{\max}, \\
		\forall n, j, m, k
	\end{array} \tag{a} \\
	& \hspace{-5em} \begin{array}{@{}l@{}}
		\|\mathbf{q}^m[n+1] - \mathbf{q}^m[n]\| \leq V^m_{\max} \Delta t, \\
		\forall n, m
	\end{array} \tag{b} \\
	& \hspace{-5em} \begin{array}{@{}l@{}}
		\mathbf{q}^m_{\text{min}} \leq \mathbf{q}^m[n] \leq \mathbf{q}^m_{\text{max}}, \\
		\forall n, m
	\end{array} \tag{c} \\
	& \hspace{-5em} \begin{array}{@{}l@{}}
		\zeta(\mathbf{q}^m[n], \mathbf{m}_j) \geq d^2(\mathbf{q}^m[n], \mathbf{m}_j) \Gamma_j^{\text{th}}, \\
		\forall j, n, m
	\end{array} \tag{d} \\
	& \hspace{-5em} \begin{array}{@{}l@{}}
		\eta \leq \text{SINR}_k, \\
		\forall k \in \mathbb{K}.
	\end{array} \tag{e}
\end{align}
	Here, \(\eta\) serves as an auxiliary variable, allowing us to convert the original Max-Min problem into a standard maximization problem. 
	This approach is particularly useful because it allows us to use standard optimization techniques to solve the problem efficiently. By maximizing 
	$\eta$, we are effectively maximizing the minimum SINR, ensuring a fair distribution of SINR among all CUs.
	This approach simplifies the problem, making it more manageable, while ensuring that the minimum SINR is maximized for all CUs.
	\subsection{Multi-Objective Formulation by Integrating Two Optimization Problems}
	\label{mf}
	In this Subsection, we introduce a novel multi-objective optimization approach by integrating two distinct optimization problems. We begin by presenting the individual optimization problems, each addressing specific objectives. The first problem (Problem \eqref{behine1}) focuses on maximizing $\Omega$ while adhering to specific constraints, while the second problem (Problem \eqref{mfi}) aims to maximize the minimum $\text{SINR}_k$ subject to its corresponding constraints. As both problems share the  common variables $\mathbf{r}_j^m [n]$ and $\mathbf{q}^m[n]$, we propose a unified multi-objective optimization framework that capitalizes on the strengths of both individual problems.
	
	
	
	To effectively address the multi-objective nature of the problem and achieve a balance between  objectives, we propose a formulation that integrates the objectives of the individual problems outlined in \eqref{behine1} and \eqref{mfi}. Specifically, we formulate an optimization problem that aims to simultaneously maximize the signal power received at the HAPS, defined as a sensing metric in \eqref{omega1}, and maximize the minimum communication SINR for CUs within  HAPS-UAV System-enabled ISAC model. The optimization involves variables such as $\mathbf{w}_k[n]$, $\mathbf{q}[n]$, $\mathbf{r}_j[n]$, and $\eta$. To identify the Pareto-optimal solutions for this multi-objective challenge, we utilize a scalarization method with a Pareto weight $\mu$ in the range $[0, 1]$, as described in \cite{boyd}:
%
%
	\begin{align}
		\label{mulll}
		\max_{\mathbf{w}_k^m [n], \mathbf{q}^m[n], \mathbf{r}_j^m [n], \eta \,\, \forall j \in \mathbb{J}, k \in \mathbb{K}} & \quad \mu \, \Omega + (1 - \mu) \, \eta \\
		\text{s.t.} \quad \quad\quad\quad\quad\quad & \hspace{-5em} \sum_{j=1}^{J} \lVert \mathbf{r}_j^m [n] \rVert^2 \leq \upsilon_m P^m_{\max}, \nonumber\\
			& \hspace{-5em}  \forall j \in \mathbb{J}, n \in \mathbb{N}, \tag{a} \\
		& \hspace{-5em} \lVert \mathbf{q}[n+1] - \mathbf{q}^m[n] \rVert \leq V_{\max} \Delta t, \nonumber\\
		& \hspace{-5em} \forall n \in \mathbb{N}, \tag{b} \\
		& \hspace{-5em} \mathbf{q}^m_{\text{min}} \leq \mathbf{q}^m[n] \leq \mathbf{q}^m_{\text{max}}, \quad 	\forall n \in \mathbb{N}, \tag{c} \\
		& \hspace{-5em} \sum_{k=1}^{K} \lVert \mathbf{w}_k^m[n] \rVert^2 + \sum_{j=1}^{J} \lVert \mathbf{r}_j^m [n] \rVert^2 \leq P^m_{\max}, \quad \nonumber \\
		& \hspace{-5em} \forall j \in \mathbb{J}, n \in \mathbb{N}, k \in \mathbb{K}, \tag{d} \\
		& \hspace{-5em} \zeta(\mathbf{q}^m[n], \mathbf{m}_j) \geq d^2(\mathbf{q}^m[n], \mathbf{m}_j) \Gamma_j^{\text{th}}, \nonumber\\
		& \hspace{-5em}  \forall j \in \mathbb{J}, n \in \mathbb{N}, \tag{e} \\
		& \hspace{-5em} \eta \leq \text{SINR}_k, \quad \forall k \in \mathbb{K}, \tag{f} \\
		& \hspace{-5em} \text{SINR}_{\text{th}} \leq \text{SINR}_k, \quad \forall k \in \mathbb{K}. \tag{g}
	\end{align}
	Here, $\mu$ is a weighting parameter that governs the relative significance of the two objectives. By adjusting $\mu$, the trade-off between maximizing $\Omega$ and maximizing $\text{SINR}_k$ can be tailored to the specific requirements of the problem. Furthermore, the multi-objective formulation encompasses the restrictions originating from both Problem \eqref{behine1} and   Problem \eqref{mfi}. Constraints in (\ref{mulll}.f) and (\ref{mulll}.g) must be satisfied for all  CUs.
	The parameter 
	$\text{SINR}_{th}$
	represents the pre-assigned SINR threshold for these CUs.

	
	
	Through the integration of the separate optimization problems, our novel multi-objective approach offers notable benefits. It provides decision-makers with the opportunity to investigate the Pareto front, highlighting a diverse array of solutions that harmonize $\Omega$ and $\text{SINR}_k$ across a range of $\mu$ values. Moreover, this method sheds light on the intricate interplay between objectives and facilitates a holistic solution that accommodates various criteria. To summarize, our introduced multi-objective framework efficiently unites two optimization challenges, leading to improved resolutions that embrace contradictory aims and attain a balanced result.
	
	
	\section{Simulation Results and Methodology Evaluation}
	\label{simi}
	\subsection{Contextual Framework and Variable Configuration}
	
	The purpose of this study is to examine the integration of HAPS  in the functioning of ISAC and UAV-based networks. To validate our proposed approach, we compare it with another strategy that involves a UAV-based ISAC network without considering HAPS,
	utilize multi-objective optimization, which serves as a widely-applied tool for improving decision-making and problem-solving within the industrial sector \cite{multiobj1}.
	
	In this Section, we present simulation results to evaluate the performance of the proposed HAPS-UAV-enabled ISAC system and gain insights into its design and implementation. In this simulation, CUs and targets are randomly placed in a square network area with dimensions of one kilometer. Additionally, various parameters used in the simulation are indicated in TABLE \ref{algorithm} unless stated otherwise.  It is assumed, that the HAPS is centrally positioned relative to all service areas. Furthermore, without loss of generality, we assume that we have a single UAV.
	The aim,  as outlined earlier, is to address the multi-objective problem \eqref{mulll} by optimizing critical variables, including the transmit beamforming vector  $\mathbf{w}_k[n]$ and $\mathbf{r}_j[n]$, and the UAV's position $\mathbf{q} [n]$, to ensure optimal system performance.

	
	\begin{table*}[t]
		\centering
		\caption{Simulation Parameters}
		\label{algorithm}
		\begin{tabular}{|c|c|c|}
			\hline
			\textbf{Parameter Name} & \textbf{Value} & \textbf{Description} \\ \hline\hline
			\multicolumn{1}{|c|}{$K$} & \multicolumn{1}{c|}{4} & \multicolumn{1}{c|}{ The number of ground CUs} \\ \hline\hline
			\multicolumn{1}{|c|}{$J$} & \multicolumn{1}{c|}{4} & \multicolumn{1}{c|}{The number of sensing-relevant targets in the area of interest} \\ \hline\hline
			\multicolumn{1}{|c|}{$G_w$} & \multicolumn{1}{c|}{ 4} & \multicolumn{1}{c|}{The number of  UAV antennas along the x-axis } \\ \hline\hline
			\multicolumn{1}{|c|}{$G_l$} & \multicolumn{1}{c|}{ 4} & \multicolumn{1}{c|}{The number of  UAV antennas along the y-axis} \\ \hline\hline
			\multicolumn{1}{|c|}{$S_w$} & \multicolumn{1}{c|}{20} & \multicolumn{1}{c|}{The number of  HAPS antennas along the width} \\ \hline\hline
			\multicolumn{1}{|c|}{$S_l$} & \multicolumn{1}{c|}{20} & \multicolumn{1}{c|}{The number of  HAPS antennas along the length} \\ \hline\hline
			\multicolumn{1}{|c|}{$\beta_{0_m} (\forall m $)} & \multicolumn{1}{c|}{-30 dB \cite{thr}} & \multicolumn{1}{c|}{The channel power gain at a reference distance $d_0 = 1\text{m}$} \\ \hline\hline
			\multicolumn{1}{|c|}{$P^m_{\max}  (\forall m)$} & \multicolumn{1}{c|}{37 dBm \cite{37dbm}} & \multicolumn{1}{c|}{Maximum transmission power from the m-th drone} \\ \hline\hline
			\multicolumn{1}{|c|}{$\sigma_{k,m}^2  (\forall m)$} & \multicolumn{1}{c|}{-110 dBm \cite{man}} & \multicolumn{1}{c|}{The noise power at each CU
				receiver} \\ \hline\hline
			\multicolumn{1}{|c|}{$d$} & \multicolumn{1}{c|}{$\lambda/2$} & \multicolumn{1}{c|}{The antenna spacing} \\ \hline\hline
			\multicolumn{1}{|c|}{$f$} & \multicolumn{1}{c|}{$120 \times 10^9$ \cite{uxnb}} & \multicolumn{1}{c|}{The carrier frequency} \\ \hline\hline
			\multicolumn{1}{|c|}{$H_m  (\forall m)$} & \multicolumn{1}{c|}{40 m\cite{thr}} & \multicolumn{1}{c|}{The  flight altitude of UAV} \\ \hline\hline
			\multicolumn{1}{|c|}{$H_{HAPS}$} & \multicolumn{1}{c|}{20000 m}&\multicolumn{1}{c|}{The  flight altitude of HAPS} \\ \hline\hline
			\multicolumn{1}{|c|}{$\Gamma^{th}$} & \multicolumn{1}{c|}{\SI{e-5}{} \cite{thr} } & \multicolumn{1}{c|}{The beampattern gain threshold} \\ \hline
	\end{tabular}
	\end{table*}
	\subsection{Optimizing HAPS-UAV-enabled ISAC  Network Using Genetic Algorithm} 
	The problem addressed in the Subsection \ref{mf} involves the challenge of finding an optimal solution for the multi-objective optimization problem presented in \eqref{mulll}. This problem is inherently non-convex, which introduces significant computational difficulties. As a result, solving such problems in practical scenarios becomes computationally intractable, classifying them as NP-hard \cite{boyd , ga33}.
	To address these complexities, we employ a metaheuristic algorithm, a class of methods well-suited for finding near-optimal solutions to difficult optimization problems. Various metaheuristic techniques, such as tabu search, particle swarm optimization (PSO), ant colony optimization, simulated annealing, and genetic algorithms (GAs), have been widely recognized for their effectiveness in tackling such challenges \cite{ga33}. 
	
	Among these, the genetic algorithm is particularly noteworthy for its adaptability, ease of implementation, and demonstrated success in optimizing CU association and UAV placement. Our approach leverages the strengths of GAs to maximize individual CU SINR while ensuring fairness throughout the system \cite{gahaps, gahaps2, ga33}.
	The Genetic Algorithm (GA) is a robust optimization method within artificial intelligence (AI), inspired by natural selection and biological evolution, that iteratively evolves solutions to tackle complex problems.
	Initially, a population of randomly selected individuals is generated from the candidate solution space. These individuals evolve towards an optimal solution over successive generations through processes such as selection, crossover, and mutation. In each generation, the fittest individuals are selected to produce a new population, thereby improving the solution quality iteratively.
	
	Given its strong global search capability, especially in complex optimization problems, the genetic algorithm is widely used in path planning, task allocation, and other similar applications \cite{gauuav, ga_2,book_ga}.
	In this paper, we propose utilizing a general-purpose heuristic search optimization tool to effectively tackle the multi-objective nature of the problem. Specifically, we employ MATLAB's "ga" algorithm, which iteratively refines a population of candidate solutions through genetic operations such as selection, crossover, and mutation. MATLAB's Optimization Toolbox provides the versatile and powerful ga function, allowing for the efficient configuration and execution of genetic algorithms. This enables the discovery of optimal or near-optimal solutions for complex problems. The specific parameters and settings used for the genetic algorithm are detailed in TABLE \ref{al2}.
	
	
	\begin{table}[t]
		\centering
		\caption{Algorithm Parameters and Settings}
		\label{al2}
		\begin{tabular}{|c|c|}
			\hline
			\textbf{Parameter} & \textbf{Value} \\
			\hline \hline
			Function tolerance & $10^{-7}$ \\ \hline \hline
			Number of population & 1700\\\hline \hline
			Crossover fraction & 0.87 \\ \hline \hline
			Mutation & Adaptive Feasible \\ \hline \hline
			Generations & 1000 \\ \hline 
			
		\end{tabular}
	\end{table}
	\subsection{Simulation Results and Performance Analysis}
	\label{simulation}
This Subsection provides a detailed analysis of the results obtained from the simulations and experiments, with a focus on evaluating the performance of the proposed HAPS-UAV system-enabled ISAC.

%
\subsubsection{Evaluation of Parameter Effects on the Proposed Framework's Performance}

\begin{figure}[t]
	\centering

\includegraphics[width=0.52\textwidth]{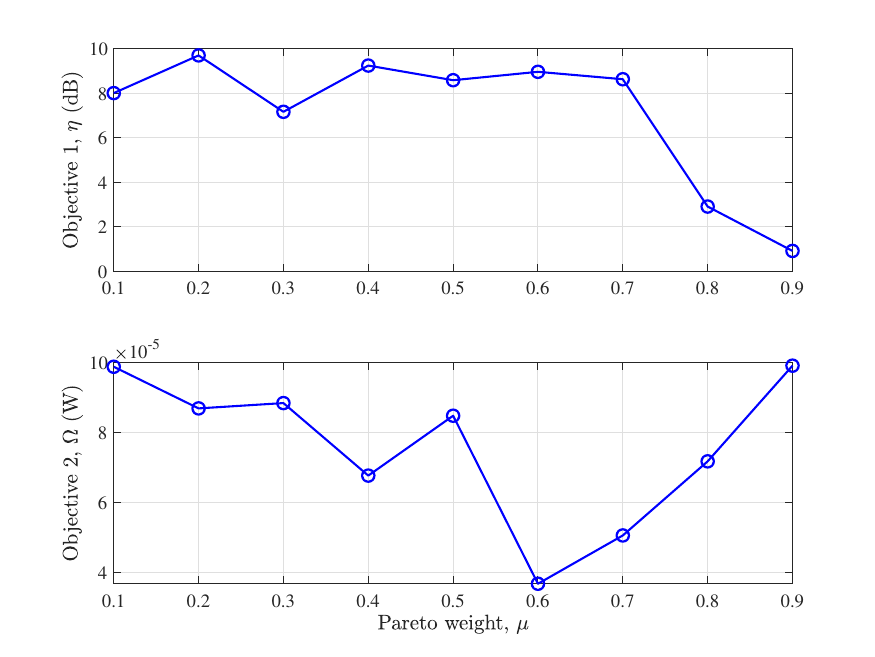}
\caption{Two-dimensional Pareto-optimal front obtained using the multiobjective genetic algorithm, illustrating the trade-off between the two objective functions as $\mu$ varies in the optimization problem defined in \eqref{mulll}.}

\label{fig1_1}
\end{figure}

Fig. \ref{fig1_1} displays the Pareto curves for different values of the Pareto weight $\mu$. The graph illustrates the trade-off between the two objective functions, $\eta$ and $\Omega$, as $\mu$ varies from 0.1 to 0.9.
By changing the value of $\mu$, whenever $\eta$ increases, $\Omega$ decreases, and vice versa. 
This behavior highlights the inherent conflict between the objectives, where an improvement in one inevitably leads to a reduction in the other. The graph clearly demonstrates how adjusting the weight $\mu$ influences the balance between $\eta$ and $\Omega$ within the context of multi-objective optimization.

\begin{figure}[t]
\centering
\includegraphics[width=0.53\textwidth]{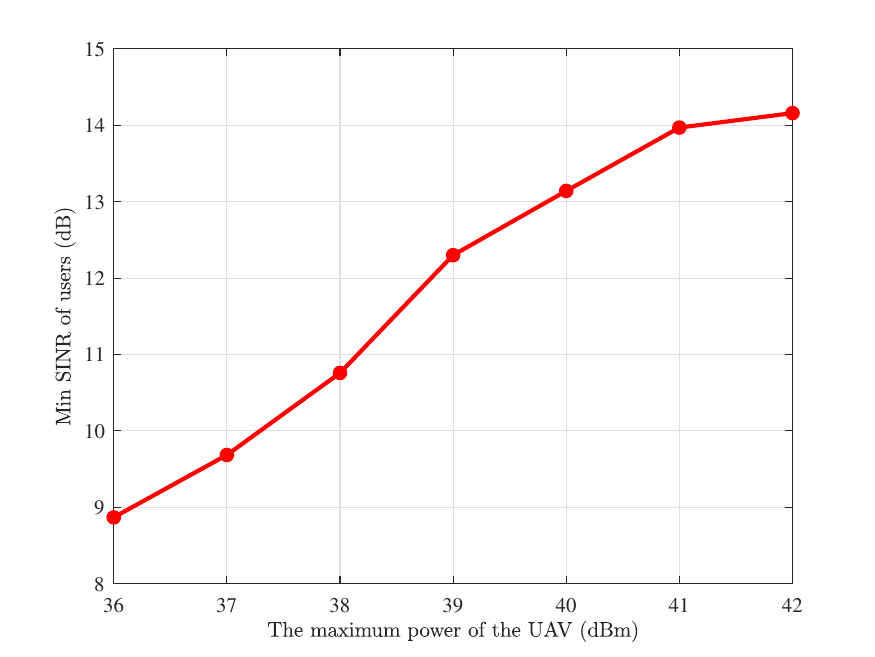}
\caption{Achievable minimum rate of CUs versus total power of UAV.}
\label{p_fig}
\end{figure}

\begin{figure}[t]
\centering
\includegraphics[width=0.53\textwidth]{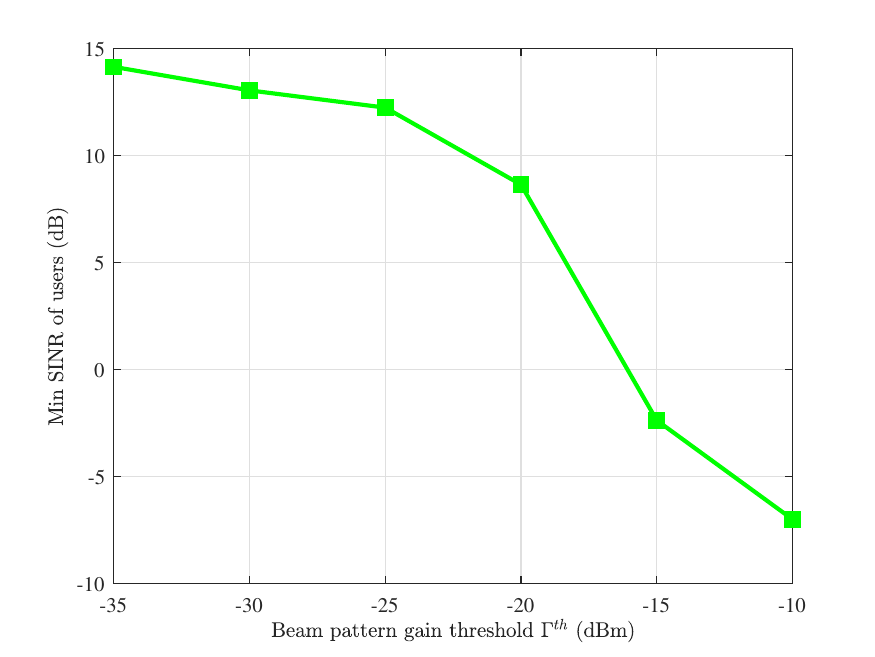}
\caption{Achievable minimum rate of CUs versus 	beam pattern gain threshold.}
\label{gama_fig}
\end{figure}
\begin{figure}[t]
\centering
\includegraphics[width=0.52\textwidth]{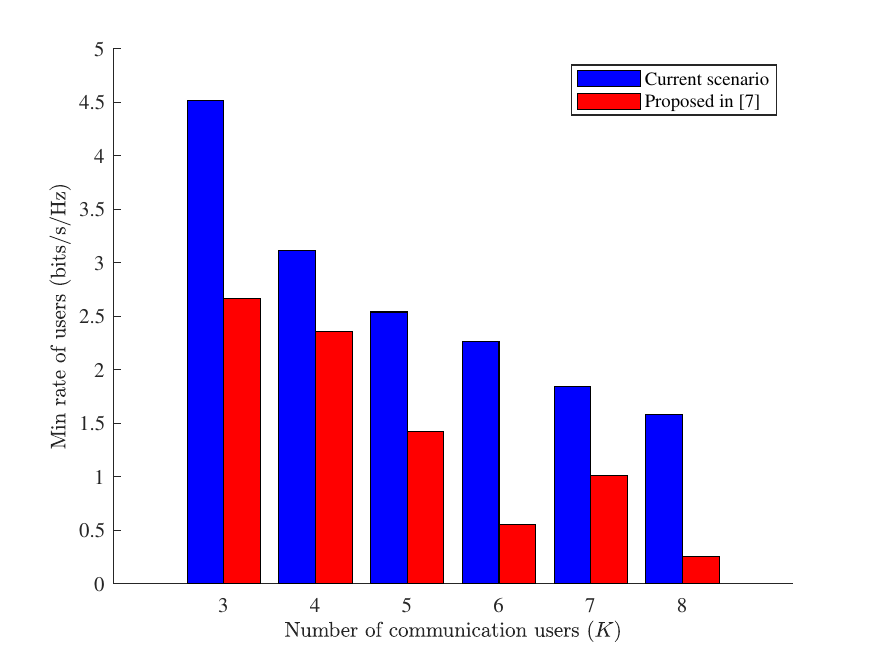}
\caption{Comparison of the rate performance of the proposed model with that of \cite{man} based on the number of communication users
	$K$. The graph shows the minimum CU rate for each value of 
	$K$ in both models.}
\label{fig2_2}
\end{figure}

\noindent
Fig. \ref{p_fig} illustrates the attainable minimum rate of CUs within the proposed HAPS-UAV System-enabled ISAC relative to the total power of  UAV.   It is evident from the figure that optimizing power allocation proves to be more beneficial for lower transmit powers, as higher transmit powers possess adequate power resources to ensure fairness among CUs.

The effect of beam pattern gain constraints on the minimum SINR of CUs is depicted in Fig. \ref{gama_fig}. As the beam pattern gain threshold ($\Gamma^{th}$) increases, there is a noticeable decline in the minimum SINR experienced by CUs. This trend highlights that more stringent beam pattern gain requirements adversely impact the SINR of the most vulnerable CUs, leading to a reduction in their achievable data rates.

\subsubsection{Comparative Analysis with Related Work}
To provide a deeper understanding, a comparative analysis is conducted with a related study. This comparison highlights the differences and similarities between the two approaches.

Fig. \ref{fig2_2} presents a comparative analysis of the minimum  CU rates
between our conducted study and the findings outlined in Reference \cite{man} across varying numbers of CUs ($K$). 
We selected \cite{man} for comparison as, based on our research, it features the system model most similar to our proposed model.
In our model, the HAPS serves as the central processor, significantly alleviating the computational burden on the UAV. This approach differs from that in \cite{man}, where the UAV solely functions as a base station, managing all signal transmissions to CUs and targets. The referenced study focuses on resource management in a network comprised exclusively of UAVs, without incorporating HAPS. As a result, the potential benefits of integrating HAPS with UAVs, which are a key feature of our model, were not explored in their work.

To maintain consistency in our analysis, we applied the genetic algorithm to solve the optimization problem in the model proposed in \cite{man}, just as we did with our own method. The genetic algorithm is a highly effective tool for addressing complex optimization challenges, especially in large, nonlinear solution spaces with multiple local optima.
In general, increasing the number of CUs often results in a decrease in the minimum CU rate, especially in resource-constrained environments. As illustrated in Fig. \ref{fig2_2}, this trend is apparent in both scenarios as the number of CUs increases.
Additionally, when comparing different CU counts, the proposed model consistently achieves a higher worst-case CU rate compared to the model referenced in \cite{man}.
This indicates a higher level of fairness in our approach.

It is worth noting that the purpose of this comparison is to observe the minimum CU rate, which reflects fairness among CUs. In future work, this comparison could be extended to other metrics, such as the sum rate.
The optimization of SINR and overall network performance using HAPS shows significant improvements over studies that focus solely on the power and flight limitations of UAVs. These findings emphasize the crucial role of ISAC systems that harness the combined strengths of both HAPS and UAVs, highlighting their superior effectiveness in enhancing network efficiency. Additionally, utilizing HAPS as the central processing unit reduces the computational load on UAVs, thereby extending their battery life.

A comparative analysis with UAV-only systems demonstrates HAPS's superior performance in resource optimization. These results clearly indicate that HAPS can effectively serve as a cost-efficient infrastructure for covering challenging areas.

\section{discussion}
\label{discusiion}
In this study, we present a thorough analysis of our optimization framework designed for integrating HAPS and UAVs within ISAC systems, specifically tailored for anticipated 6G networks. This discussion emphasizes the unique advantages offered by the HAPS-UAV synergy compared to existing methodologies, particularly those focusing exclusively on UAV-centric ISAC systems.
\subsection{Algorithms} 
To address the optimization challenges inherent in our framework, we employed a combination of genetic algorithm (GA) for multi-objective optimization. The selection of GA is due to its effectiveness in navigating non-convex problem spaces commonly encountered in resource allocation scenarios, allowing us to achieve near-optimal solutions efficiently. The flexibility of GA facilitates the adjustment of multiple parameters while comprehensively treating the interactions between communication and sensing requirements. In contrast to classical methods like Depth-First Search, which may overlook complex trade-offs in dynamic environments, GA provides a more adaptable and refined solution strategy.
\subsection{Methods of Solution}
\noindent Our proposed solution incorporates several key techniques:\\
\noindent \textbullet\hspace{0.5em} Multi-objective Optimization: We integrated two distinct optimization problems—maximizing the minimum SINR
for  CUs and maximizing the signal power received at the HAPS. This dual approach ensures balanced resource allocation that effectively meets both communication and sensing needs.\\
\noindent \textbullet\hspace{0.5em} Dynamic Power Allocation: The power allocation strategy is designed to fulfill stringent sensing performance metrics without compromising communication capacity. By adapting power usage according to the specific demands of each CU and target, we optimize network efficiency and extend UAV operational ranges.\\
\noindent \textbullet\hspace{0.5em} Beamforming Techniques: The implementation of a Uniform Planar Array (UPA) for directed signal transmission showcases advanced beamforming capabilities. This strategy enhances transmission to maximize signal integrity and quality, contrasting sharply with earlier methodologies that inadequately addressed the unique attributes of HAPS.

\subsection{Capabilities} 
Our analysis reveals several performance enhancements compared to existing solutions:
\begin{itemize}
\item Improved Signal Quality: The integration of HAPS into the communication framework led to a significant increase in the minimum SINR for ground CUs. This enhancement not only meets but exceeds communication requirements while effectively managing concurrent sensing tasks, a capability not demonstrated in previous UAV-centric models.
\item Reduced Latency and Resource Consumption: By offloading computational tasks to HAPS, we alleviate the energy constraints faced by UAVs, resulting in longer operational durations in high-demand scenarios. Our framework's ability to streamline processing through HAPS has been shown to decrease latency and improve data handling efficiencies in dynamic environments.
\item Adaptability: The flexible resource allocation strategies employed allow for rapid adaptation within the HAPS-UAV ecosystem, unlike the more static methodologies seen in prior studies. This flexibility is essential for addressing the nuanced demands of real-time communication and sensing operations, particularly in complex urban terrains.
\end{itemize}

\subsection{Comparative Analysis} 
When compared to existing UAV-centric ISAC frameworks, our approach proves more robust in handling the intricacies of simultaneous communication and sensing. While traditional models tend to focus on UAVs in isolation, our integration of HAPS creates a complementary relationship that enhances overall system performance. This synergy not only improves resource utilization but also addresses common constraints associated with UAV operations, such as limited bandwidth and higher latency.
\subsection{Limitations and Future Directions}
While our findings are promising, challenges remain, particularly in real-world applications where factors like signal interference, obstacle occlusions, and variances in environmental conditions can affect system performance. Future research should explore adaptive modulation strategies for UAV-HAPS interactions under varying operational conditions, enhancing overall robustness and efficiency.
\subsection{Overall Assessment}
Our study establishes that the integration of HAPS into ISAC paradigms significantly enhances the performance, flexibility, and robustness of communication infrastructures crucial for the evolving landscape of 6G networks. The ability of HAPS to offload computational tasks not only improves operational efficiency but also broadens application scopes, particularly in remote areas. This foundational work paves the way for further investigations into multi-platform integration dynamics, highlighting the potential of cooperative HAPS and UAV systems in next-generation wireless communication frameworks.

\section{conclusion}
\label{natije}
This article explores the integration of HAPS and UAVs within an ISAC framework, with a focus on its relevance for 6G wireless networks. Utilizing a multi-objective optimization approach with genetic algorithms, we have enhanced network performance through the HAPS-UAV ISAC system. HAPS and UAVs are identified as critical components for extending wireless coverage, especially in remote and challenging environments. HAPS technology offers continuous coverage, cost-effective infrastructure, and increased network capacity, making it essential for future 6G systems. By designating HAPS as the CPU, the proposed model alleviates the computational burden on UAVs, thus preserving their energy resources and improving network performance. Offloading computational tasks to HAPS not only boosts UAV efficiency but also extends their operational lifespan, contributing to a more robust network infrastructure. 

Simulation results validate the effectiveness of our approach in optimizing resource allocation and ensuring fairness across the network. Our approach demonstrates significant improvements in fairness and system capabilities compared to traditional methods, particularly when benchmarked against the reference study. These results underscore the potential of integrating HAPS, UAVs, and ISAC systems to enhance network performance, communication efficiency, and support diverse applications in wireless networks.
\bibliography{rafmain4}
\bibliographystyle{ieeetr}

\end{document}